\documentclass[runningheads]{llncs}
\usepackage[T1]{fontenc}
\usepackage[linesnumbered,ruled,vlined]{algorithm2e}
\usepackage{graphicx}
\usepackage[sort,nocompress]{cite}
\usepackage{xcolor}
\usepackage{wrapfig}
\usepackage{amsmath} 
\usepackage{caption}
\usepackage{booktabs}   % \toprule \midrule \bottomrule
\usepackage{makecell}   % \makecell
\usepackage{caption}    % \captionof / \captionsetup
\usepackage{multirow}
\usepackage[hidelinks]{hyperref}
\usepackage{orcidlink}
\usepackage{bbding}
\newcommand{\revise}[1]{\textcolor[rgb]{0.00,0.00,0.00}{#1}}
\newcommand{\TODO}[1]{\textcolor[rgb]{0.00,0.00,0.00}{#1}}
\begin{document}
\title{Realizable N:M Sparse Transformer Inference via Search–Kernel Co-Design}

\author{
Yiming Liu\inst{1}\orcidlink{0009-0009-2079-8359} \and
Wenqi Lou\inst{2}\Envelope\orcidlink{0000-0002-2240-6672} \and
Zhiguang Wang\inst{3}\orcidlink{0009-0002-5829-3339} \and
Zhiwei Ke\inst{1}\orcidlink{0009-0003-7636-2446} \and \\
Fengrui Zuo\inst{1}\orcidlink{0009-0003-3468-9280} \and
Chao Wang\inst{1}\orcidlink{0000-0002-9403-5575} \and
Xuehai Zhou\inst{1}\orcidlink{0000-0002-8360-3143}
}

\authorrunning{Y. Liu et al.}

\institute{
University of Science and Technology of China, Hefei, China
\and
Suzhou Institute of Advanced Research, University of Science and Technology of China, Suzhou, China
\and
National Key Laboratory of Modeling and Simulation for Complex Systems, China\\
\email{liuyiming9974@mail.ustc.edu.cn}, \email{louwenqi@ustc.edu.cn}, \email{cswang@ustc.edu.cn}
}

\maketitle              % typeset the header of the contribution

\begin{abstract}
% Vision Transformers (ViTs) achieve strong accuracy but suffer from high inference latency. Semi-structured N:M sparsity is attractive because it reduces arithmetic cost, yet its theoretical savings often fail to translate into proportional end-to-end speedup on modern GPUs. The key challenge is that deployment latency depends not only on arithmetic reduction but also on execution regularity and hardware scheduling under sparsity. Realizable acceleration, therefore, requires coordinated design across both sparse execution and sparsity configuration.
Vision Transformers (ViTs) achieve strong accuracy but incur high inference latency. Semi-structured N:M sparsity can reduce arithmetic cost, yet its theoretical savings often fail to translate into proportional end-to-end speedups on modern GPUs. This mismatch arises because deployment latency depends not only on arithmetic reduction but also on execution regularity and hardware scheduling under sparsity. Achieving practical acceleration, therefore, requires coordinated design across sparse execution and sparsity configuration.
To this end, we propose a hardware-software co-design framework for N:M sparse ViT inference. On the hardware side, we design MD-SpMM, an N:M sparse CUDA kernel that reorganizes sparse GEMM into micro-dense, Tensor-Core-aligned dataflow and uses inference-aware adaptive parallelism to sustain utilization. 
% On the software side, we perform layer-wise sparsity search under explicit end-to-end latency budgets, so that sparsity allocation is guided by deployment behavior rather than abstract sparsity objectives. 
% Experiments on multiple ViT/Swin models and GPU platforms show that, under target speedup constraints, the proposed co-design achieves better accuracy--latency trade-offs with predictable and realizable performance gains.
\revise{
On the software side, we perform layer-wise sparsity search under explicit end-to-end latency budgets using a three-stage heuristic search with constraint relaxation to avoid premature convergence and enable deployment-aware sparsity allocation.
Experiments on multiple ViT/Swin models and GPU platforms show that the framework achieves over 2.2× latency speedup while maintaining comparable accuracy and delivering superior accuracy under the same latency constraint. The source code is publicly available at \url{https://github.com/liuganhuo/realizable-nm-sparse-transformer}.}

\end{abstract}

\section{Introduction}
Vision Transformers (ViTs) leverage self-attention to capture long-range dependencies and achieve strong performance in tasks such as image classification and semantic segmentation~\cite{deit}. However, this benefit comes with high inference cost: ViTs rely heavily on matrix multiplications in projection and feed-forward layers, imposing substantial compute and memory demands that make deployment on real hardware increasingly challenging~\cite{vitcod,glvlsi24_qin}.

Model compression, especially pruning, has been widely studied to mitigate this burden by removing redundant parameters while preserving accuracy~\cite{pieee_compression}. Among pruning schemes, semi-structured N:M sparsity has emerged as a practical compromise between accuracy preservation and hardware regularity, and has been adopted across CNNs, ViTs, and related accelerators~\cite{iclr21_nmsparse_cnn,nmkernel,hpca22_s2ta,aspdac25_nm}.

Despite this progress, the theoretical savings of N:M sparsity often fail to translate into proportional end-to-end latency gains in real inference, especially under INT8 deployment, as evidenced by the experimental comparisons \TODO{in Section~6}. 
In this regime, latency depends not only on arithmetic reduction but also on execution regularity, memory behavior, and scheduling effects under sparsity. This reveals a persistent gap between theoretical efficiency and practical performance. The gap is often misinterpreted as sparsity being ``ineffective'', whereas it is more accurately attributed to the difficulty of realizing sparsity efficiently on current deployment stacks.

We therefore focus on \emph{realizable acceleration}: whether N:M sparsity can deliver a target end-to-end latency reduction on a real deployment stack, rather than merely improving proxy metrics such as FLOPs, global sparsity, or isolated operator speedups~\cite{icml22_spdy,yang2018netadapt,sun2021dominosearch}. From a system perspective, realizable acceleration requires satisfying two tightly coupled conditions. First, sparse computation must be \emph{execution-realizable}: sparse operators should deliver stable and efficient execution on real hardware instead of exhibiting irregular behavior that prevents consistent latency reduction~\cite{nmkernel}. Second, sparsity assignment must be \emph{latency-grounded}: when N:M sparsity is applied in a layer-wise manner, per-layer ratios should be selected under explicit deployment-level latency constraints rather than abstract sparsity or FLOPs-based objectives that poorly reflect end-to-end cost~\cite{cvpr24_elsa}. Violating either condition leads to unrealizable speedups in practice, even when theoretical computation reduction is significant. Yet existing approaches often optimize only one side~\cite{he2018amc,aspdac25_nm,nmkernel,lou2024unleashing}, motivating an end-to-end co-design framework that couples execution-realizable sparse kernels with latency-grounded layer-wise configuration and lightweight candidate evaluation.

To this end, we propose a hardware-software co-design framework for realizable N:M sparse inference acceleration. On the hardware side, we design MD-SpMM, an N:M sparse CUDA kernel that restructures sparse computation into Tensor-Core-aligned execution through micro-dense restructuring and adaptive parallelism. 
% On the software side, we introduce a latency-grounded layer-wise search pipeline that combines a recoverability-driven zero-shot proxy, a profiling-based latency predictor, and a hardware-aware evolutionary solver to evaluate candidates without repeated fine-tuning. 
\revise{On the software side, we introduce a latency-grounded layer-wise search pipeline that combines a profiling-based latency predictor and a feasible-region-aware evolutionary solver to evaluate candidates efficiently.}
By integrating execution-realizable kernels with latency-grounded sparsity search, the framework consistently translates sparsity-induced computation reduction into end-to-end latency improvement on real deployment stacks.

In summary, this work makes the following contributions:

\begin{itemize}

    \item We present a \textbf{search--kernel co-design framework} for \emph{realizable} N:M sparse Transformer acceleration, built on execution realizability and configuration realizability under deployment latency constraints, and enabling more accurate and reliable acceleration under the same target speedup.

    \item We design \textbf{MD-SpMM}, an execution-realizable Tensor-Core-native N:M sparse kernel that converts sparse computation into regular MMA-centric execution with scalable inference parallelism. Across diverse matrix multiplication shapes and sparsity levels, it delivers 2.0$\times$ average speedup over cuBLAS dense GEMM and 1.6$\times$--4.0$\times$ speedup over \revise{nmSPARSE~\cite{nmkernel}}.
    
    % \item We propose a \textbf{latency-grounded layer-wise N:M search pipeline} combining a recoverability-driven zero-shot proxy, a calibrated LUT-based latency predictor, and a feasible-region-aware evolutionary solver. It enables training-free, deployment-aware search, achieving {270$\times$} faster search than prior work while delivering better latency--accuracy trade-offs under the same target speedup across ViT/Swin inference settings.
    \item{
    \revise{We propose a latency-grounded layer-wise N:M search pipeline that integrates a calibrated LUT-based latency predictor with a feasible-region-aware evolutionary solver to efficiently discover high-quality sparse configurations. With our hardware–software co-design, the resulting solutions achieve over 2.2× latency reduction while delivering higher accuracy, yielding up to +1.4\% Top-1 accuracy on ImageNet-1K and consistently outperforming prior methods across ViT and Swin inference settings.}
    }
\end{itemize}
\section{Background and Motivation}

\subsection{Vision Transformers and Layer-wise N:M Configuration}
Vision Transformers (ViTs) process an image as a sequence of patch embeddings and forward it through a stack of Transformer blocks. In practice, inference is dominated by the Q/K/V projections, the attention output projection, and the two FFN linear layers. For common vision workloads, these matrix-heavy operators account for most of the end-to-end inference time, making them the primary targets for sparsification~\cite{vitcod}.

\begin{figure}[t]
    \centering
    \begin{minipage}[t]{0.5\textwidth}
        \centering
        \includegraphics[width=\linewidth]{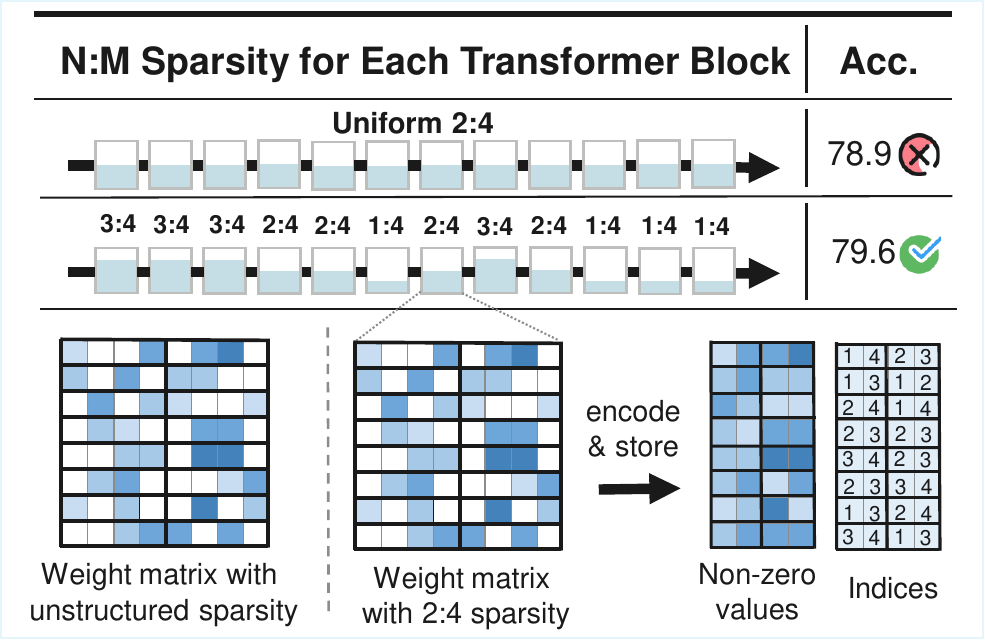}
        \caption{N:M sparsity and uniform vs.\ layer-wise configurations in ViTs.}
        \label{fig1}
    \end{minipage}
    \hfill
    \begin{minipage}[t]{0.47\textwidth}
        \centering
        \includegraphics[width=\linewidth]{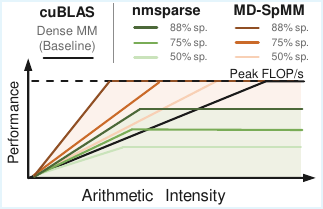}
        \caption{Roofline analysis of N:M sparse INT8 inference on GPUs.}
        \label{fig2}
    \end{minipage}
\end{figure}

N:M sparsity constrains each group of $M$ weights to retain $N$ nonzeros. In ViTs, it is often applied uniformly across all Transformer blocks, as illustrated in Fig.~\ref{fig1}. However, despite their similar block structure, layers at different depths contribute unequally to model accuracy and inference latency due to their depth-dependent roles~\cite{cvpr24_elsa,sun2021dominosearch,fu2025unicos}. This leads to heterogeneous sparsity sensitivity across blocks, making a single global N:M ratio suboptimal. Consequently, ViT sparsification is more naturally formulated as a layer-wise N:M configuration problem.

\subsection{The Lack of Efficient Sparse Execution Mechanisms}

Although N:M sparsity reduces multiply-accumulate operations, it does not
directly yield efficient execution on modern Tensor Core GPUs. This mismatch is especially pronounced for INT8 inference, where dense GEMM already achieves high efficiency through regular dataflow, predictable control flow, and high MMA utilization~\cite{nvidia_a100,mishra2021accelerating,chen2021efficient}. In this regime, sparse speedup depends not only on arithmetic reduction but also on whether execution remains compatible with the hardware execution model. Existing sparse kernels often rely on metadata handling, runtime decoding, and irregular memory accesses, which disrupt Tensor Core dataflow and reduce utilization~\cite{gale2020sparse,nmkernel,wang2025unisparta}.

As illustrated by the Roofline analysis in Fig.~\ref{fig2}, existing approaches exhibit a clear execution trade-off. nmSPARSE~\cite{nmkernel} follows sparse decoding and irregular access patterns, leading to low arithmetic intensity and limited utilization, whereas cuBLAS sustains high efficiency with dense GEMM but cannot exploit N:M structure. This gap highlights the need for sparse execution mechanisms that preserve Tensor-Core-friendly regularity while exposing the arithmetic benefit of N:M sparsity.

\subsection{The Need for Latency-Grounded Layer-wise Configuration}

Introducing latency constraints into sparse configuration search is necessary but challenging, since meeting a target latency requires different sparsity levels across models, layers, and deployment environments. Traditional methods that optimize proxy metrics such as sparsity ratio or FLOPs often fail to capture real deployment behavior. Even with similar overall sparsity, different layer-wise assignments can lead to substantially different end-to-end latencies. Moreover, enforcing latency constraints fragments the feasible space by discarding invalid configurations, weakening search-space connectivity and making exploration prone to narrow feasible regions.

To address this issue, sparse configuration optimization must incorporate deployment-level latency semantics. In particular, latency models built on the target hardware and deployment stack are needed to guide the search toward configurations that both satisfy latency constraints and enable realizable end-to-end acceleration. This mitigates the mismatch between abstract optimization objectives and real hardware behavior~\cite{litepred,icml22_spdy,tang2026closertome}.

\section{Problem Formulation and Method Overview}

Given a Transformer model and a target end-to-end latency budget, the problem is to realize N:M sparse inference that preserves model quality while delivering actual speedup on the target hardware.

\begin{figure*}[t]
    \centering
    \includegraphics[width=\textwidth]{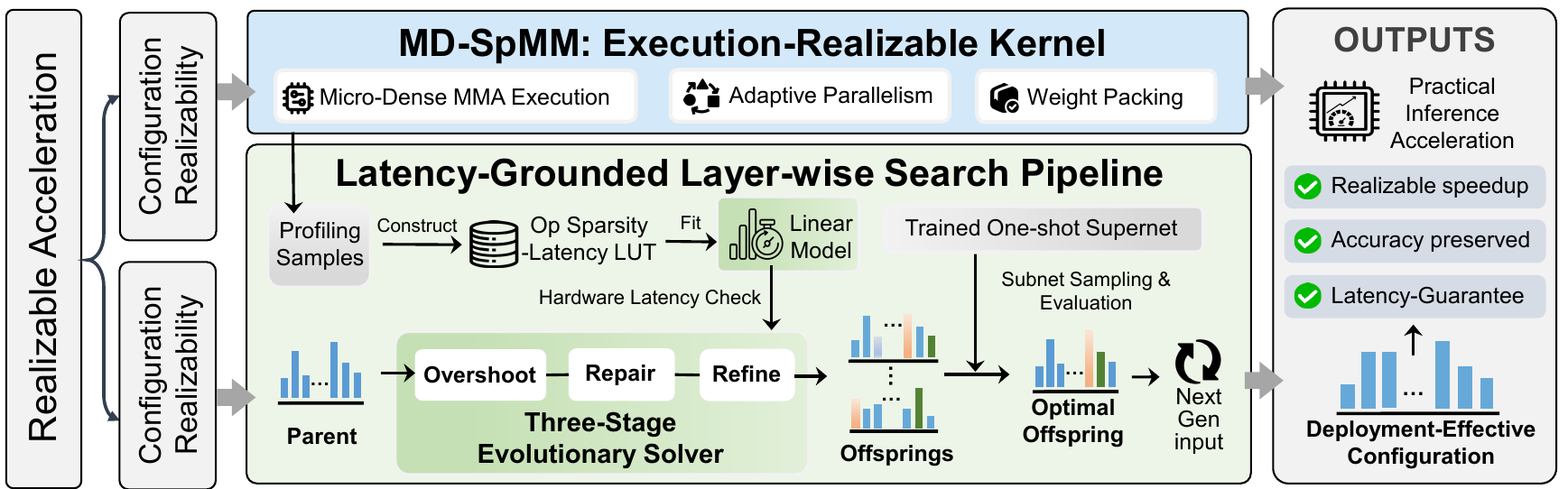}
    \caption{\TODO{Overview of the search-kernel co-design framework for N:M sparse Transformer acceleration.}}
    \vspace{-2pt}
    \label{fig:framework}
\end{figure*}

As shown in Fig.~\ref{fig:framework}, we address this problem through two coordinated components. At the execution level, we design MD-SpMM, a sparse kernel that realizes N:M sparsity through Tensor-Core-aligned execution. At the configuration level, we build a latency-grounded layer-wise search pipeline, in which a calibrated LUT-based latency predictor enforces deployment feasibility and a hardware-aware search strategy explores the fragmented feasible region under explicit latency constraints.

Together, these components instantiate the two requirements of realizable acceleration introduced in Section~1. MD-SpMM addresses execution realizability, while the latency-grounded search addresses configuration realizability by identifying layer-wise N:M assignments aligned with end-to-end deployment behavior. In this sense, realizable acceleration is achieved only when sparse execution and sparsity configuration are jointly optimized.

\section{MD-SpMM: Dataflow-Centric N:M Sparse Kernels}

\subsection{Design Methodology}

From the perspective of execution realizability, INT8 inference with N:M sparsity depends not only on reduced arithmetic but also on whether sparse execution matches the hardware execution model. We identify two first-order requirements: (i) \emph{dataflow regularity} to sustain Tensor Core MMA utilization, and (ii) \emph{inference-scale parallelism} to maintain high SM occupancy across varying operator shapes. Violating either requirement by introducing index-driven control, irregular memory access, or insufficient thread-block parallelism can erase theoretical sparsity gains in end-to-end latency. As shown in Fig.~4, MD-SpMM uses three mechanisms: (1) Weight Packing for decode-friendly sparse storage, (2) Micro-Dense Execution for MMA-centric regular tiles, and (3) Adaptive Parallelism to improve occupancy while bounding reduction overhead.

\subsection{Micro-Dense: From N:M Sparsity to Tensor Core Dataflow}
To make N:M sparsity compatible with Tensor Core execution, we introduce \emph{Micro-Dense}, which reformulates N:M SpMM as an MMA-centric regular dataflow. Under N:M sparsity, each window of size $M$ contains $N$ nonzeros. Rather than exposing them to the compute stage as irregular sparse operands, Micro-Dense materializes only these $N$ nonzeros into fixed MMA-compatible dense tiles aligned with Tensor Core fragment shapes, so Tensor Cores still operate on regular dense fragments and runtime computation remains entirely MMA-based.

This design shifts sparsity irregularity out of the MMA compute path and into the loading-and-decoding stage. To keep the resulting overhead low, we adopt a decode-friendly packed representation (\texttt{B\_pack}) that jointly encodes nonzero values and their positions within each N:M window. This removes separate index loads and enables lightweight decoding, which is overlapped with shared-memory placement and MMA issuance.

Therefore, Micro-Dense confines sparsity handling to packed loading and lightweight decoding, allowing runtime computation to operate directly on MMA-aligned tiles. Overall, it establishes a clear execution mapping: N:M sparsity $\rightarrow$ Micro-Dense tiles $\rightarrow$ Tensor Core dataflow.
\begin{figure*}[t]
    \centering
    \includegraphics[width=\textwidth]{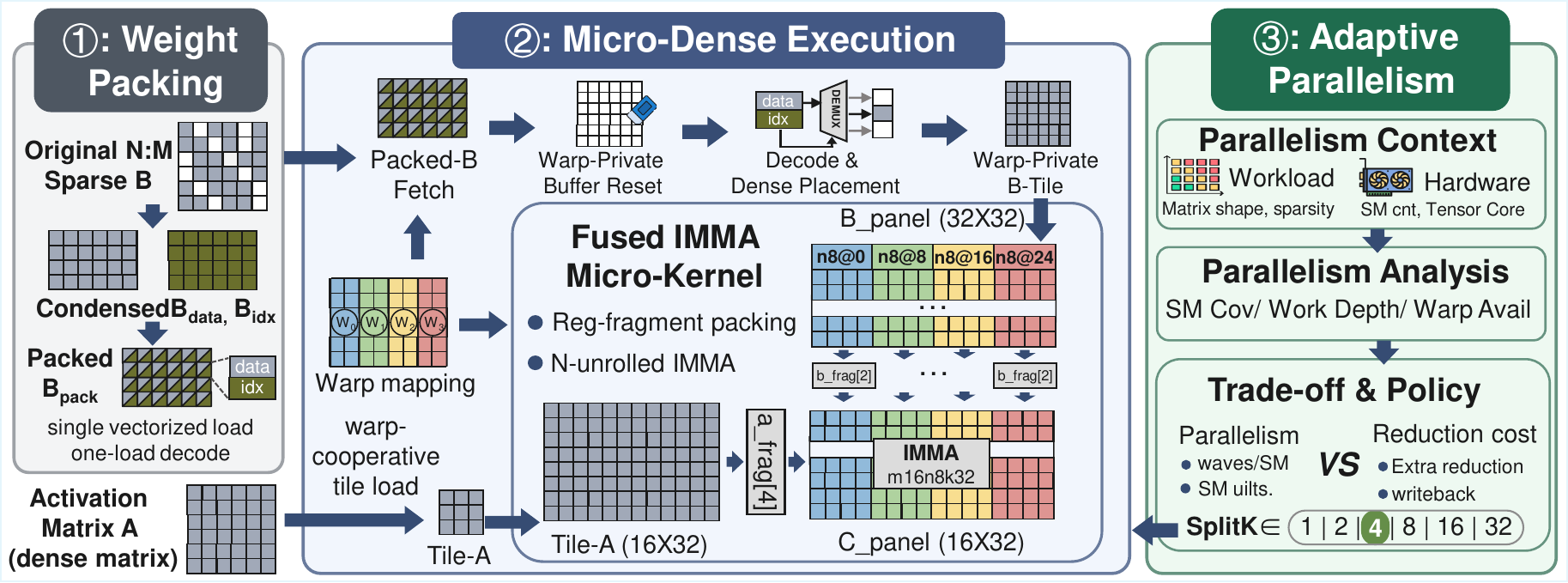}
    \caption{Overview of the MD-SpMM kernel design.}
    \vspace{-2pt}
    \label{fig:kernel}
\end{figure*}

\subsection{Adaptive Parallelism: Split-K for Inference-Scale SpMM}

Micro-Dense improves per-block Tensor Core efficiency, but inference can still suffer from \emph{under-occupancy} when operator shapes expose limited parallel work. End-to-end performance therefore depends not only on efficient single-block execution, but also on scalable and predictable parallelism across workloads.

We address this with split-K parallelism, which partitions the reduction dimension so multiple thread blocks compute partial sums in parallel, expanding the parallel space from the output dimensions $(M, N)$ to $(M, N, \textit{splitK})$. Split-K boosts occupancy but introduces reduction and extra write-back overhead, so its benefit must be controlled.

We therefore adopt a rule-driven splitK selection strategy. First, \textbf{parallelism priority}: splitK is increased until the number of blocks is sufficient to fully schedule SMs. Second, \textbf{controlled reduction}: splitK is restricted to a small discrete candidate set so that K-splits remain aligned with Tensor Core granularity and reduction overhead stays predictable. When $K$ is small or reduction dominates, the strategy reverts to a smaller splitK. 

In this approach, all splits share the same compute template and partial sums are merged only in the final stage, avoiding additional control complexity. Combined with Micro-Dense, this adaptive parallelism enables consistent speedup across deployment scenarios.

\section{Latency-Grounded Layer-wise Sparsity Search}

% \TODO{While MD-SpMM makes N:M sparsity execution-realizable, selecting an effective layer-wise configuration remains challenging due to expensive candidate evaluation and deployment-dependent latency behavior. We therefore build a latency-grounded search pipeline that combines a recoverability-driven zero-shot proxy, a LUT-based latency predictor, and a hardware-aware search strategy.}

While MD-SpMM makes N:M sparsity execution-realizable, determining an effective layer-wise sparsity configuration remains challenging, since accuracy alone is insufficient and the end-to-end latency impact of sparsity is strongly deployment-dependent. We therefore develop a latency-grounded search strategy that explicitly incorporates latency constraints to guide the search toward higher-quality and deployment-feasible configurations.

\subsection{Lookup Table-based Latency Modeling}

Accuracy ranking alone is insufficient, since the search must also satisfy an end-to-end latency budget. Because measuring full-model latency for every candidate is too costly, we build a lightweight predictor for feasibility checks~\cite{litepred,fu2025unicos,weihong,wang2025unicox}.

\begin{figure}[t]
    \centering
    \includegraphics[width=0.98\linewidth]{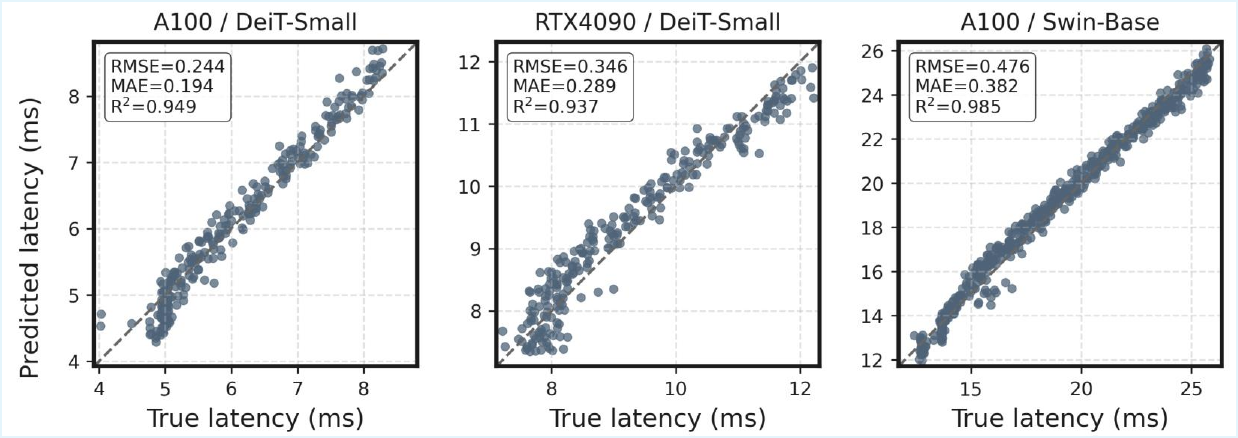}
    \caption{Predicted vs.\ measured end-to-end latency on platforms and architectures.}
    \label{fig:fit}
    \vspace{-0.4cm}
\end{figure}

Under a fixed hardware--software stack, we profile MD-SpMM operators and build an operator-level sparsity--latency lookup table (LUT). For each operator shape and candidate N:M ratio, we benchmark sparse matrix multiplication latency (average repeated runs). Empirically, the latency is largely determined by operator shape and sparsity level, yielding a LUT indexed by $(\textit{shape}, \textit{sparsity})$. Since ViT-family models reuse only a small set of operator shapes, the LUT can be rebuilt offline within seconds when the deployment environment changes.

% \vspace{-0.8cm}
To predict end-to-end latency, we fit a first-order linear model on top of the LUT. Let $t_{\ell}(s_{\ell})$ denote the LUT latency of layer $\ell$ under sparsity setting $s_{\ell}$, then
\begin{equation}
T_{\text{pred}}=\sum_{\ell=1}^{L} \gamma_{\ell}\, t_{\ell}(s_{\ell}) + b,
\end{equation}
where $\gamma_{\ell}\ge 0$ and $b\ge 0$ are calibrated by non-negative least squares (NNLS) using a small number of full-model measurements. 
This calibration absorbs graph-level scheduling, runtime overhead, and compiler optimizations beyond single-operator profiling. After calibration, the predictor enables constant-time feasibility checks during search with within-5\% error (Fig.~\ref{fig:fit}), grounding layer-wise sparsity decisions in deployment latency rather than abstract sparsity metrics.

\subsection{Hardware-Aware Layer-wise Sparsity Search}
\begin{algorithm}[t]
	\small
	\caption{\textbf{HW-aware heuristic search algorithm}}
    \label{alg:search}
	\label{Alg_Global}
        \KwIn{Model $M$; latency budget $\mathcal{T}_{\text{target}}$; number of generations $\mathcal{G}$; number of offspring $\lambda$;\textcolor{gray}{ //define: parent $x$; child $y$}.}
        $\textbf{Initialization:}$
        $candidate = \mathtt{UniSparGen}(\mathcal{T}_{target}, M);$\\
        $x_{1} = \mathtt{Rand\_Spar}(candidate,\, toward=\mathcal{T}_{target});$\\

        \textbf{Search: }\For{ $g$  \textbf{in} $[1, \mathcal{G}]$}{
             $offsprings = [];$\\
            \textbf{3Stage-OffspringGen: }\For{ $i$ \textbf{in} $[1,\; \lambda]$}{
                $y_{i} = x_{g};$\\
                \For{$l$ \textbf{in} $\mathtt{Random\_K}(\,\{\, l\, |\, y_{i}[l]<8\, \})$\,}{
    		    $y_{i}[l]=N+1;$ \textcolor{gray}{//Densify the $l$-th layer of $y_i$ (e.g., 4:8 to 5:8), keeping other layers unchanged.}}
                \While {$Latency(y_{i}) > \mathcal{T}_{target}$}{
                    $l = \mathtt{Random\_1}(\,\{\, l\, |\, y_{i}[l] >1 \,\}\,);$\\
                    $y_{i}[l] = N-1;$ \textcolor{gray}{//Sparsify the $l$-th layer of $y_i$}\\
                }
                $D := \{\, l \mid (y_i[l]<8) \wedge\ \Delta\mathtt{Fitness}_{l} \ge 0\};$ \textcolor{gray}{//layers whose densification yields a fitness gain.}\\
                \While {$D \neq \emptyset$}{
                    $y_{i}[\mathtt{Random\_1}(D)]\!=\!N\!+\!1;$ \textcolor{gray}{//Random densify one layer}\\
                    $D := \{\, l \mid (y_i[l]<8) \wedge\ \Delta\mathtt{Fitness}_{l} \ge 0\};$ \textcolor{gray}{//Update}\\
                }
                $offsprings.append (y_{i});$\\
            }
             \textbf{Selection: } $x_{g+1} = \mathtt{MAX\_Fitness}(offsprings);$\\
        }
    \Return $x_{\mathcal{G}+1};$ \text{(final sparse configuration)}
\end{algorithm}

Given the latency predictor, we search for a layer-wise N:M configuration that maximizes model quality under a target latency budget. Let $M$ denote the model, and let $s=[s_1,\dots,s_L]$ denote the layer-wise N:M configuration, where each $s_{\ell}$ is selected from a discrete set of hardware-supported N:M ratios. We solve
\begin{equation}
\max_{s} \; Acc\big(\mathcal{M}(s)\big)
\quad
\text{s.t.} \quad
\mathcal{T}_{\text{pred}}\big(\mathcal{M}(s)\big)\le \mathcal{T}_{\text{target}},
\end{equation}
where $Acc(\cdot)$ denotes the Top-1 accuracy on ImageNet-1K and $\mathcal{T}_{\text{target}}$ is the target end-to-end latency budget.

This problem is challenging because, under layer-wise N:M sparsity, the mapping from sparsity configuration to deployment latency is highly nonlinear. The latency constraint therefore partitions the search space into multiple disconnected feasible regions. As a result, prior local-perturbation-based methods~\cite{sieberling2024evopress} tend to get trapped on small ``feasibility islands'', causing the search to stagnate.

To address this, we adopt a feasible-region-aware $(1+\lambda)$ evolutionary search that follows an \emph{overshoot--repair--refine} procedure, enabling effective transitions across fragmented feasible regions in the sparsity--latency landscape. The overall pipeline is summarized in Algorithm~\ref{alg:search}.
The search is instantiated within the standard one-shot sparsity search framework~\cite{cvpr24_elsa}, where the supernet training stage remains unchanged and each candidate configuration is scored by its inherited accuracy under the trained supernet. 

During initialization, we first construct several latency-feasible configurations with approximately uniform sparsity distributions and select the one with the highest $\mathtt{fitness}$ score (Eq.~\ref{eq:fitness}) as the initial parent. We then repeatedly invoke \textbf{3Stage-OffspringGen} to produce $\lambda$ offspring via the following three stages:

\begin{equation}
    \label{eq:fitness}
    \mathtt{Fitness}(\mathcal{S}) =
    \begin{cases}
    \mathtt{Acc}_{\text{Inherit}}\big(\mathcal{M}(s)\big), & \mathcal{T}_{\text{pred}}\big(\mathcal{M}(s)\big) \le T_{\text{target}}, \\
    -\infty, & \text{otherwise}.
    \end{cases}
\end{equation}

1) \textit{Overshoot} (Lines 7--8) adopts a ``relax-first'' strategy: a small random subset of layers is moved to the next denser sparsity level, deliberately violating the latency budget. This avoids the vanishing effect (noise-level score variations) of conventional 1:1 ``sparsify-dens'' perturbations near the feasibility boundary.

2) \textit{Repair} (Lines 9--11) then restores feasibility by increasing sparsity on randomly sampled layers until the latency constraint is met, avoiding the directional bias of greedy, latency-guided repair and encouraging diverse search trajectories.

3) \textit{Refine} (Lines 12--15) then explores the feasible boundary by gradually densifying the model while retaining only latency-valid updates. Candidate configurations with non-decreasing $\mathtt{fitness}$ are accepted, allowing the search to advance along the feasible boundary iteratively.

Throughout the search, the latency predictor provides constant-time feasibility checks, and the final configuration is selected through offspring competition under the latency constraint. Overall, this design enables more effective exploration of the fragmented feasible space and improves the quality of deployment-feasible layer-wise N:M configurations.
\section{Experiments}

\subsection{MD-SpMM Kernel Evaluation}\label{sec:kernel_eval}
% \textbf{Benchmark}
% We evaluate MD-SpMM using a benchmark across matrix shapes and sparsity levels. The matrices are defined by the dimensions $M$, $N$, and $K$ for matrix multiplication, where $M$ is the number of rows, and $N$ and $K$ are the number of columns in the input and output matrices, respectively. We consider sparsity levels of 50\%, 75\%, and 87.5\%, with $M \in \{64, 128, 256, 512\}$ and $N, K \in \{128, 256, 512, 1024\}$. Experiments are performed on A100, RTX 4060, and RTX 4090 GPUs using Ubuntu 22.04 and CUDA 12.2, with C++/CUDA kernel implementation compiled using \texttt{nvcc}.

\textbf{Benchmark}
We evaluate MD-SpMM using a benchmark across matrix shapes and sparsity levels. The matrices are defined by the dimensions $M$, $N$, and $K$ for matrix multiplication, where $M$ is the number of rows, and $N$ and $K$ are the number of columns in the input and output matrices, respectively. We consider sparsity levels of 50\%, 75\%, and 87.5\%, with $M \in \{64, 128, 256, 512\}$ and $N, K \in \{128, 256, 512, 1024\}$. Experiments are performed on A100, RTX 4060, and RTX 4090 GPUs using Ubuntu 22.04 and CUDA 12.2, with C++/CUDA kernel implementation compiled using \texttt{nvcc}. We report latency excluding one-time offline weight packing, since \texttt{B\_pack} construction is performed once before deployment. All runtime costs during inference are included, including decoding overhead, data movement, and MMA computation. Thus, both kernel latency and end-to-end latency correspond to pure inference-time execution.

\textbf{Baselines}
We compare MD-SpMM against state-of-the-art dense and sparse baselines: cuBLAS, NVIDIA’s highly optimized dense GEMM library representing the vendor-optimized upper bound, and nmSPARSE~\cite{nmkernel}, a representative library for N:M sparse matrix multiplication.

% \textbf{Results}
% Figure~\ref{fig:exp1} reports the kernel speedup of MD-SpMM and nmSPARSE over dense cuBLAS across matrix shapes under different N:M sparsity levels. The vertical axis shows speedup normalized to dense cuBLAS, while the horizontal axis indexes different matrix shapes; speedups above 4 are truncated for clarity. 

% MD-SpMM consistently outperforms nmSPARSE and delivers stable acceleration across matrix shapes and sparsity levels. By contrast, nmSPARSE provides only limited or moderate gains in many cases, indicating that irregular sparse execution often offsets the theoretical benefit of N:M sparsity. 
% These results show that practical N:M sparse acceleration requires not only arithmetic reduction, but also regularized Tensor-Core-friendly execution and inference-scale-aware parallelism. Overall, MD-SpMM substantially improves the execution realizability of N:M sparsity under deployment-relevant conditions.
\begin{table}[t]
    \centering
    \caption{Adaptive parallel strategy across platforms and sparsity levels.}
    \label{tab:adaptive_parallel}
    \setlength{\tabcolsep}{4pt}
    \small
    \resizebox{\linewidth}{!}{
    \begin{tabular}{c|ccc|ccc|ccc}
        \toprule
        Platform &
        \multicolumn{3}{c|}{A100} &
        \multicolumn{3}{c|}{RTX 4090} &
        \multicolumn{3}{c}{RTX 4060} \\
        \cmidrule(lr){1-1} \cmidrule(lr){2-4} \cmidrule(lr){5-7} \cmidrule(lr){8-10}
        Sparsity &
        50\% & 75\% & 87.5\% &
        50\% & 75\% & 87.5\% &
        50\% & 75\% & 87.5\% \\
        \midrule
        Top-1 Hit Rate    & 0.813 & 0.875 & 0.827 & 0.799 & 0.711 & 0.690 & 0.720 & 0.763 & 0.714 \\
        Avg. Norm. Perf.  & 0.990 & 0.990 & 0.990 & 0.983 & 0.965 & 0.958 & 0.972 & 0.968 & 0.960 \\
        Worst Norm. Perf. & 0.818 & 0.774 & 0.803 & 0.786 & 0.746 & 0.739 & 0.773 & 0.694 & 0.696 \\
        \bottomrule
    \end{tabular}}
    \vspace{-0.4cm}
\end{table}
\begin{figure*}[t]
    \centering
    \includegraphics[width=\textwidth]{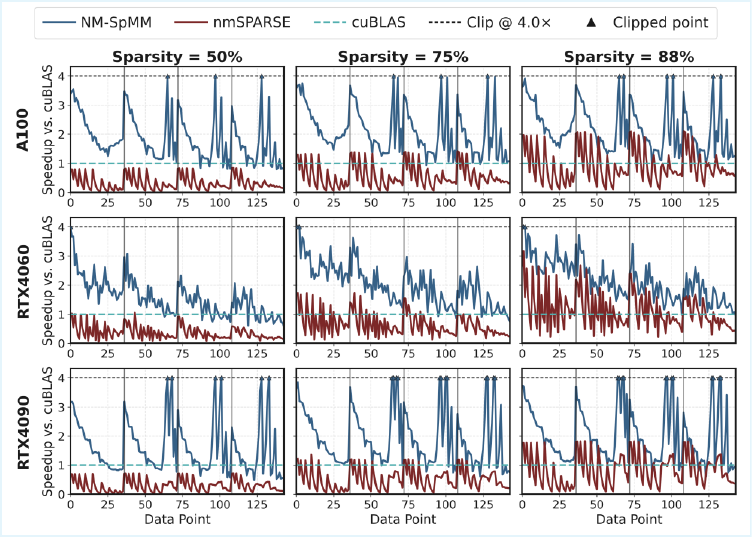}
    \vspace{-0.4cm}
    \caption{Speedup over cuBLAS across matrix shapes under different sparsity levels.}
    \label{fig:exp1}
    \vspace{-0.4cm}
\end{figure*}

\textbf{Results}
We first evaluate the adaptive parallel strategy against fixed Split-K settings. For each shape--sparsity pair, we test multiple Split-K candidates and use the lowest latency as the oracle optimum. Table~\ref{tab:adaptive_parallel} summarizes the Top-1 hit rate, average normalized performance, and worst-case normalized performance. The adaptive strategy selects the oracle-optimal configuration in most cases and remains close to optimal even in non-hit cases, while also maintaining strong worst-case robustness. Compared with a fixed parallel setting, it improves overall kernel performance by more than 50\%.

Fig.~\ref{fig:exp1} reports the kernel speedup of MD-SpMM and nmSPARSE over dense cuBLAS across matrix shapes under different N:M sparsity levels. The vertical axis shows speedup normalized to dense cuBLAS, while the horizontal axis indexes different matrix shapes; speedups above 4 are truncated for clarity.

 MD-SpMM consistently outperforms nmSPARSE and delivers stable acceleration across platforms. On A100, it achieves around 2.0$\times$ average speedup over cuBLAS and 1.6$\times$--4.0$\times$ speedup over nmSPARSE~\cite{nmkernel}. Similar trends are observed on RTX 4060 and RTX 4090, with absolute gains varying due to hardware characteristics such as compute capability and memory bandwidth.

These results show that practical N:M sparse acceleration requires not only arithmetic reduction, but also regularized Tensor-Core-friendly execution together with shape-aware parallel adaptation. Overall, MD-SpMM substantially improves the execution realizability of N:M sparsity under deployment-relevant conditions.

\subsection{\revise{Feasible-Region-Aware Search
Validation}}\label{sec:proxy_search_eval}

\begin{figure*}[t]
    \centering
    \includegraphics[width=\textwidth]{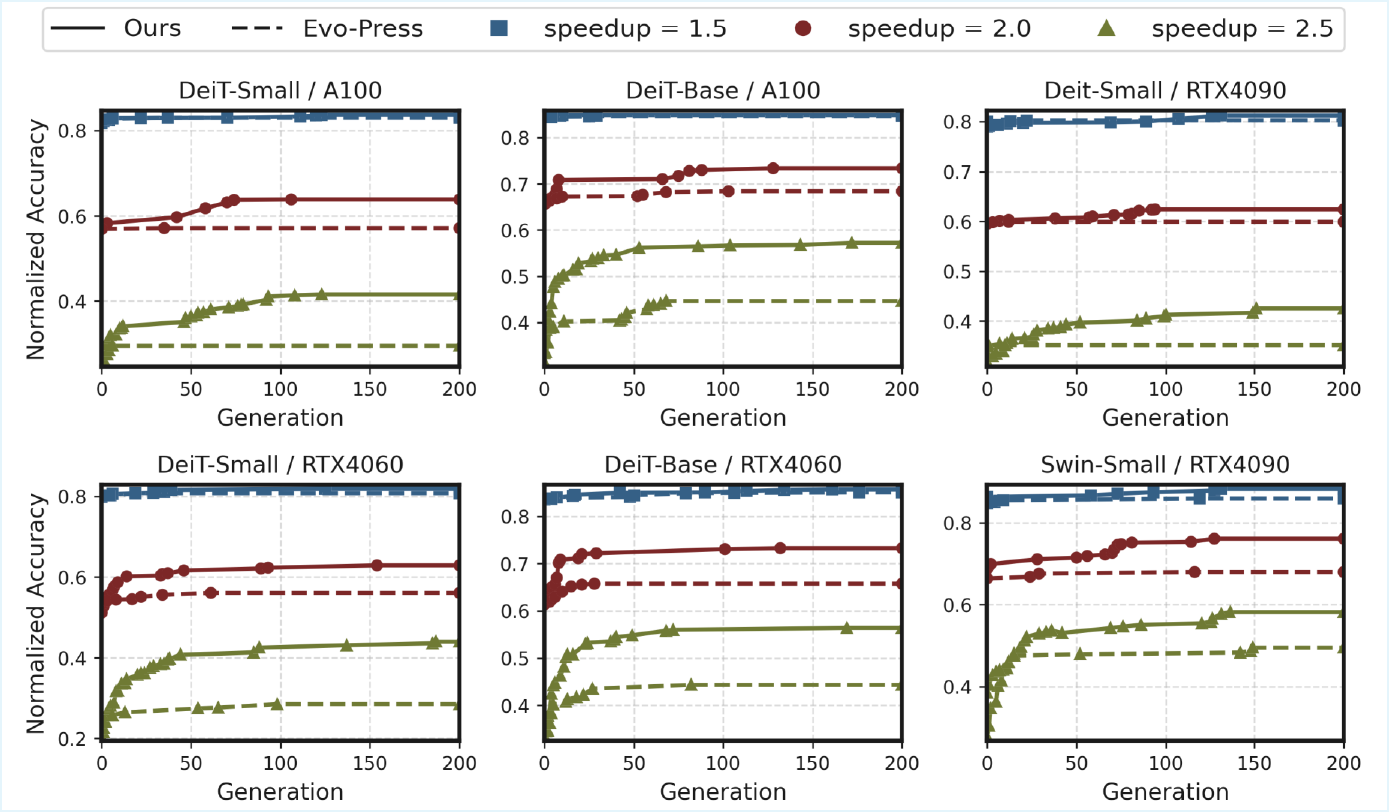}
    \vspace{-0.4cm}
    \caption{\TODO{Comparison of search trajectories between Evo-Press and our method across models and speedup settings.}}
    \label{fig:search}
    \vspace{-0.8cm}
\end{figure*}

\textbf{Benchmark:}
We evaluate the proposed search strategy on DeiT-Small, DeiT-Base, and Swin-Small under 1.5$\times$, 2.0$\times$, and 2.5$\times$ target speedups on A100 , RTX 4060 and RTX 4090 GPUs. To improve efficiency, search is conducted on ImageNet-300, a subset of ImageNet-1K with 300 classes, 500 training images, and 50 validation images per class. We find that it preserves the search trends and final configurations observed on full ImageNet-1K while substantially reducing search cost.

\textbf{Baseline:}
We instantiate our method within the standard one-shot layer-wise sparsity search framework, keeping supernet training unchanged and replacing only the post-training search stage with our latency-grounded search algorithm. We compare against Evo-Press~\cite{sieberling2024evopress}, a representative evolutionary method for layer-wise sparsity search, under the same one-shot framework and latency prediction setup. Under this setting, the dominant cost still comes from supernet training; the full pipeline with 200 search generations takes about 10 hours on three A100 GPUs.

\textbf{Results:}
Fig.~\ref{fig:search} shows the search trajectories of our method and Evo-Press across different models and target speedups. Our method exhibits more stable convergence and consistently reaches better final solutions, while Evo-Press more often stagnates near inferior local regions. In many cases, our method overtakes Evo-Press as early as the 20th--30th generation, and the gap further widens under higher speedup targets, where the feasible region becomes more constrained and fragmented. Moreover, the search typically converges within about 100 generations, indicating good search efficiency in practice. These results demonstrate that explicitly incorporating latency constraints into the search process improves both search quality and robustness under practical deployment budgets.

\subsection{Overall Evaluation}\label{sec:overall_eval}

\textbf{Benchmark}
To evaluate the end-to-end effectiveness of the proposed search-kernel co-design framework, we conduct INT8 sparse inference experiments on DeiT-Small and DeiT-Base using an NVIDIA A100 GPU. All methods are tested under the same inference stack and deployment setting. 

\textbf{Baselines}
We compare different combinations of sparsity configuration strategies and execution backends. On the search side, we consider uniform N:M sparsity, EvoPress~\cite{sieberling2024evopress}, and our latency-grounded layer-wise search. On the execution side, we compare nmSPARSE~\cite{nmkernel} and MD-SpMM as sparse execution backends, while cuBLAS serves as the dense reference. This allows us to isolate the effect of search, kernel, and their combination under the same deployment constraint.

\begin{figure*}[t]
    \centering
    \includegraphics[width=\textwidth]{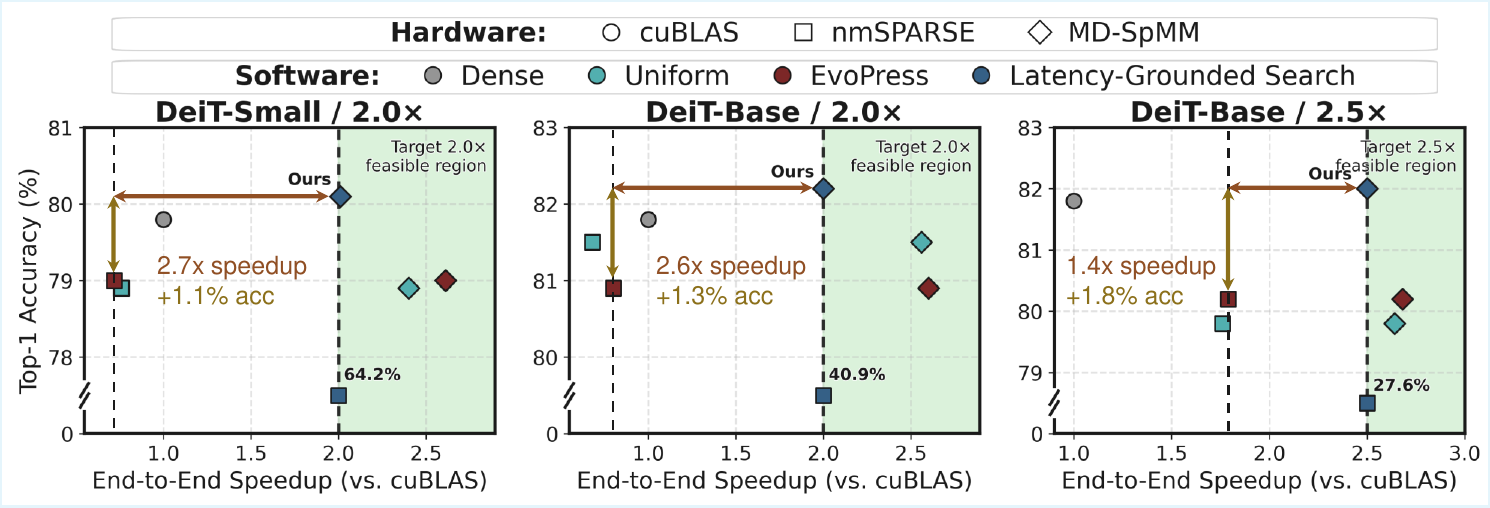}
    \caption{Accuracy-speedup trade-offs across search and kernel combinations.}
    \label{fig:overall}
    \vspace{-0.8cm}
\end{figure*}

\textbf{Results:}
Fig.~\ref{fig:overall} summarizes the resulting accuracy--speedup trade-offs, where the shaded feasible region denotes configurations satisfying the target speedup. Improving sparse execution alone is insufficient if sparsity allocation remains weak or latency-unaware. When deploying N:M sparse models with nmSPARSE, irregular execution overhead often offsets arithmetic savings, so more aggressive sparsity is required to meet the target speedup, which in turn causes noticeable accuracy degradation. Conversely, improving sparsity allocation alone is also insufficient without a realizable sparse backend. \TODO{Uniform N:M sparsity and proxy-driven approaches, such as EvoPress, optimize abstract objectives that are misaligned with deployment latency, often over-pruning latency-critical layers or producing configurations that fail to satisfy the latency constraint.} Together, these results show that practical sparse acceleration requires both realizable sparse execution and latency-aligned layer-wise sparsity search.

% By contrast, the proposed framework achieves the best overall result by coupling latency-grounded search with MD-SpMM: the former allocates sparsity according to deployment behavior, while the latter ensures these allocations translate into realizable end-to-end speedup. As a result, it attains higher accuracy within the feasible region and satisfies the latency target more reliably than methods that optimize only search or only execution.

% Overall, these results show that realizable N:M acceleration benefits most from jointly optimizing layer-wise sparsity configuration and sparse execution.

By contrast, the proposed framework achieves the best overall result by coupling latency-grounded search with MD-SpMM. The former allocates layer-wise sparsity according to deployment behavior, while the latter ensures that these allocations translate into realizable end-to-end speedup on hardware. As a result, compared with conventional strategy combinations, our method consistently delivers better accuracy--speedup trade-offs across all evaluated settings, with an average 2.2$\times$ speedup gain and 1.4\% accuracy improvement.

Overall, these results highlight a key principle for practical N:M acceleration: neither sparsity search nor sparse execution alone is sufficient. Favorable accuracy--speedup trade-offs require jointly optimizing layer-wise sparsity configuration and hardware-realizable sparse execution, so that algorithmic sparsity decisions remain aligned with deployment performance.

\section{Conclusion}
This paper argues that practical N:M sparse Transformer acceleration is a search-kernel co-design problem. MD-SpMM makes sparse execution more realizable on Tensor-Core GPUs, while the proposed latency-grounded layer-wise search pipeline allocates sparsity by deployment behavior rather than abstract sparsity objectives. Together, they turn N:M sparsity into a more stable and deployment-effective acceleration mechanism under explicit latency targets.

\begin{credits}
\subsubsection{\ackname}
This work was supported in part by the Strategic Priority Research Program of the Chinese Academy of Sciences, Grant Nos. XDB0660101, XDB0660000, and XDB0660100, in part by the National Natural Science Foundation of China under Grant No. 62502489, in part by Jiangsu Provincial Natural Science Foundation under Grants BK20241818 and BK20250479.
\subsubsection{\discintname}
The authors have no competing interests to declare that are relevant to the content of this article.
\end{credits}

\bibliographystyle{splncs04}
\bibliography{ref}





% LNCS Homepage, \url{http://www.springer.com/lncs}, last accessed 2023/10/25
% \end{thebibliography}
\end{document}